\documentclass[aip, amsmath,amssymb,preprint]{revtex4-1}

\usepackage{graphicx}
\usepackage{dcolumn}
\usepackage{bm}

\usepackage[utf8]{inputenc}
\usepackage[T1]{fontenc}
\usepackage{mathptmx}
\usepackage{etoolbox}

\makeatletter
\def\@email#1#2{%
 \endgroup
 \patchcmd{\titleblock@produce}
  {\frontmatter@RRAPformat}
  {\frontmatter@RRAPformat{\produce@RRAP{*#1\href{mailto:#2}{#2}}}\frontmatter@RRAPformat}
  {}{}
}%
\makeatother
\begin{document}

\preprint{}

\title[Inherent Non-Linear Damping in Resonators with Inertia Amplification]{Inherent Non-Linear Damping in Resonators with Inertia Amplification}
\author{B. Van Damme} 
\email{bart.vandamme@empa.ch.}
\author{G. Hannema}%
\author{L. Sales Souza}%
\author{B. Weisse}
\author{D. Tallarico}%
\author{A. Bergamini}%

\affiliation{ 
Empa, Materials Science and Technology\\
Ueberlandstrasse 129, 8600 Duebendorf\\
Switzerland}%

\date{\today}

\begin{abstract}
Inertia amplification is a mechanism coupling degrees of freedom within a vibrating structure. Its goal is to achieve an apparent high dynamic mass and, accordingly, a low resonance frequency. Such structures have been described for use in locally resonant metamaterials and phononic crystals to lower the starting frequency of a band gap without adding mass to the system. This study shows that any non-linear kinematic coupling between translational or rotational vibrations leads to the appearance of amplitude-dependent damping. The analytical derivation of the equation of motion of a resonator with inertia amplification creates insight in the damping process, and shows that the vibration damping increases with its amplitude. The theoretical study is validated by experimental evidence from two types of inertia-amplification resonators. Finally, the importance of amplitude-dependent damping is illustrated when the structure is used as a tuned mass damper for a cantilever beam.
\end{abstract}

\maketitle

Metamaterials are structures with a desired homogenized macroscopic property that is induced by their meso-scale composition~\cite{deymier2013acoustic}. Static properties such as Young's modulus and Poisson ratio can be achieved that significantly differ from those of the base material~\cite{bertoldi2017flexible}. By choosing an appropriate mesoscopic scale, optical~\cite{cai2010optical}, acoustic~\cite{cummer2016controlling}, and elastic~\cite{bertoldi2017flexible} wave processes can be altered at will. Some notable desired effects are cloaking~\cite{landy2013full,hao2010super}, sound absorption~\cite{jimenez2016ultra}, vibration isolation~\cite{bergamini2019tacticity}, or wave guiding~\cite{miniaci2018experimental}. Most successful designs, which already found their way to applications, rely on efficient models of wave dispersion within periodic media~\cite{de2019dynamic}. Bloch-Floquet boundary conditions can be exploited to predict the location and efficiency of band gaps~\cite{mace2008modelling}. Topological wave guides are based on more exotic dispersive effects, such as (double) Dirac cones~\cite{huber2016topological}. Similar exotic wave phenomena, such as band gaps and localization, can be achieved by aperiodic arrangements, in which case extensive numerical models have to replace reduced unit cell models with periodic boundaries~\cite{colombi2016seismic,van2021bending}. 

The unit cell approach is valid for linear structures, in which case the wave speed does not depend on the excitation amplitude. Elastic structures with a non-linear stress-strain relation can only be accurately described by their dispersion relation to a certain extent, e.g. for a harmonic excitation at a certain amplitude~\cite{khajehtourian2014dispersion,bukhari2020spectro}. The propagation of a general signal, containing multiple frequencies at random amplitudes, cannot be accurately predicted by the dispersion alone~\cite{van1997quasi,xiang2007modulation}. Well-known wave effects in non-linear metamaterials, such as solitons in granular chains~\cite{nesterenko1984propagation} or amplitude gaps in periodic structures with geometric non-linearities~\cite{deng2018metamaterials}, can only exist if the different frequency components have amplitude-dependent speeds. The dynamics of these structures have to be derived from the explicit equations of motion.

One of the main quests in vibration isolation is finding suitable materials that are at the same time stiff, light, and attenuating. Locally resonant metamaterials add tuned resonators to a vibrating host structure to reduce its vibration amplitude at desired frequencies~\cite{liu2000locally}. Attenuating low frequencies requires higher masses or soft connectors, which is not acceptable for many engineering applications. Alternatively, shunted piezo resonators have been used as a promising lightweight solution~\cite{flores2017bandgap}. It is however possible to reduce the resonance frequency of purely mechanical structures by coupling degrees of freedom in two and three dimensions. The structure's motion in the vibration direction leads to the excitation of additional inertial elements, either translational (point masses)~\cite{yilmaz2007phononic} or rotational (disks)~\cite{bergamini2019tacticity}. The coupled motion results in an apparent increase of mass, leading to a much lower eigenfrequency using a minor increase in weight and without loss of stiffness. The efficiency of this inertia amplification (IA) mechanism has been proven in various setups, well predicted by linearized methods. 

\begin{figure}
    \centering
    \includegraphics[width = \columnwidth]{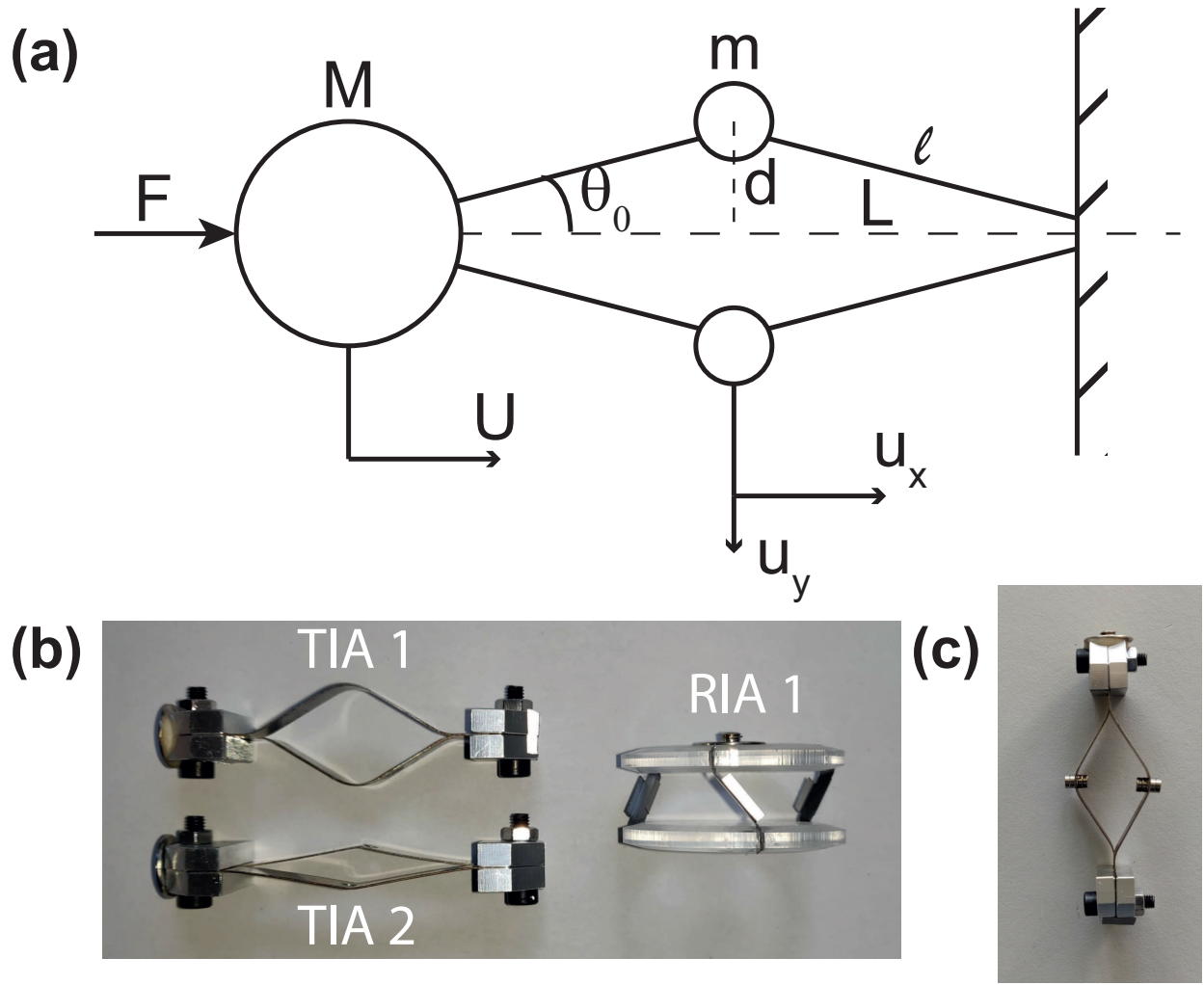}
    \caption{(a) Schematic representation of a resonator with inertia amplification. The motion of the main mass $M$ is coupled to the amplification masses $m$. (b) Experimental samples to investigate the nonlinear kinematic damping: two with translational amplification (TIA), one with rotational amplification (RIA). (c) IA structure with 4 magnetic point masses on each beam.}
    \label{fig:samples}
\end{figure}

In this Letter, we look into the dynamics of individual resonators with IA, showing that the motion coupling leads to inherent non-linear damping. This can be shown from an IA structure coupling the translational motion of the main mass $M$ with the masses $m$ (Fig.~\ref{fig:samples}(a)). A horizontal displacement $U$ induces the translation of the auxiliary mass $u_x = U/2$ and $u_y = \sqrt{\ell^2 - (L - U/2)^2} - d$. The associated velocities can be found by taking the derivative with respect to time: 
\begin{eqnarray}
    \dot{u}_x &=& \dot{U}/2\\
    \dot{u}_y &=& \frac{L - \frac{U}{2}}{2 \sqrt{\frac{L^2}{\cos^2\theta_0} - \left(L - \frac{U}{2}\right)^2}} \dot{U},\label{eq:veloc}
\end{eqnarray}
where $\ell = L/\cos \theta_0$.
The velocity can be expanded as a Taylor series for small strains $\epsilon = U/2L$:
\begin{equation}
    \dot{u}_y = \dot{U}\left[ \frac{1}{2 \tan \theta_0} + \frac{\epsilon}{2 \tan\theta_0 \sin^2\theta_0} + \frac{3}{4}\frac{\epsilon^2}{\tan^3\theta_0 \sin^2\theta_0}\right].
\end{equation}
The 0th order term proportional to $1/\tan\theta_0$ is the linearized amplification term, increasing drastically for low values of $\theta_0$. However, if this angle is sufficiently small, the higher order terms do not vanish for realistic strains, and truncating the Taylor series leads to large errors. This shows that highly amplified systems behave inherently nonlinear.

The dynamics of a system with coupled degrees of freedom can, in general, be elegantly derived using the Lagrangian $\mathcal{L} = T - V$. For the system with translational IA (TIA) shown in Fig.~\ref{fig:samples}(a), $T$ is the kinetic energy $T = M|\dot{U}|^2/2 + m|\dot{u}|^2$ and $V$ is the potential energy $V = k U^2/2$. The resonator's spring constant $k$ depends on the connectors between the masses. For this special case, where the generalized force and coordinate are equal to the physical force $F$ and coordinate $U$, the equation of motion for the mass $M$ is defined by 
\begin{equation}
    F = \frac{d}{dt}\left( \frac{\partial \mathcal{L}}{\partial \dot{U}}  \right) - \frac{\partial \mathcal{L}}{\partial U}.
\end{equation}
This yields
\begin{equation}\label{eq:eom}
    F = \tilde{M} \ddot{U} + \tilde{C} \dot{U} |\dot{U}| + \tilde{K} U, 
\end{equation}
with 
\begin{eqnarray}
\tilde{M} &=& M + m\left[ 1 + \frac{(1-\epsilon)^2}{A} \right]\\
\tilde{C} &=& \frac{m}{2L} \left[ \frac{1-\epsilon}{A}  + \frac{(1-\epsilon)^3}{A^2}\right]\\
\tilde{K} &=& k \\
A &=& \frac{1}{\cos^2 \theta_0} - (1-\epsilon)^2.
\end{eqnarray}

To gather insight in these rather intricate equations we assume that $\theta_0$ is not too small, so that the 0-th order term of the Taylor expansion can be used for small strains:
\begin{equation}
\tilde{M} = M + \frac{m}{\sin^2\theta_0} \qquad \tilde{C} = \frac{m}{2L} \frac{\cos^2\theta_0}{\sin^4\theta_0}.
\end{equation}
The apparent added mass is equal to $m/\sin^2\theta_0$, in other words the masses are dynamically increased by an amplification factor $\alpha = 1/\sin^2\theta_0$. In~\cite{bergamini2019tacticity}, an asymptotic model for a disk supported by flexural beams has shown that such amplification is only valid at non-vanishing angles, otherwise non-physical results are obtained. More interestingly, the derivation shows that the non-linear kinematic relation between $U$ and $u_y$ yields a nonlinear damping term, which depends on the amplification mass, and which increases for smaller angles $\theta_0$. It would seem tempting to use the smallest possible angles to achieve a low eigenfrequency $\omega_0 = \sqrt{\tilde{K}/\tilde{M}}$ with a minimal auxiliary mass. However, the increasing damping yields an unpractically high value for $\tilde{C}$, and consequentially an overdamped system. 

A second interesting feature of the equation of motion is the non-linear nature of the damping force, proportional to the square of the velocity. The absolute value results mathematically from the fact that the kinetic energy is positive for positive and negative signs of $\dot{U}$ in the Lagrange equation, and physically ensures that IA leads to an amplitude attenuation in the absence of external forces. This particular form of the damping term is known from drag forces, and is therefore well described in literature~\cite{worden2001nonlinearity}. Assuming a harmonic force $F = F_0 \sin(\omega t - \phi)$, the steady-state solution can be approximated by $U = U_0 \sin(\omega t)$. Plugging these relations into equation~(\ref{eq:eom}) leads to the well known form of the frequency response function $\Lambda$, the response amplitude normalized by the force amplitude for a one degree-of-freedom system, 
\begin{equation}
    \Lambda =\frac{U_0}{F_0} =  \frac{1}{\tilde{K} +  i c_{eq}\omega -\tilde{M}\omega^2},
\end{equation}
with
\begin{equation}
     c_{eq} = \frac{8\tilde{C} \omega U_0}{\pi}.
\end{equation}
In the weakly-nonlinear case of small displacement amplitudes, the harmonic distortion is negligible, so that we can assume that the velocity amplitude for a harmonic response is given by $V_0 = \omega U_0$, and thus 
\begin{equation}
     c_{eq} = \frac{8\tilde{C} V_0}{\pi}.
\end{equation}

Structures with rotational IA (RIA) behave in a similar way, but the kinetic energy is amplified by the rotation of a disk. The resulting formulas for apparent mass and damping coefficients are very alike~\cite{suppmat}. We can conclude that TIA and RIA resonators behave like a damped mass-spring system. The damping coefficient depends on the auxiliary mass and connector angle, and increases with the excitation amplitude. 

\begin{figure}
    \centering
    \includegraphics[width = \columnwidth]{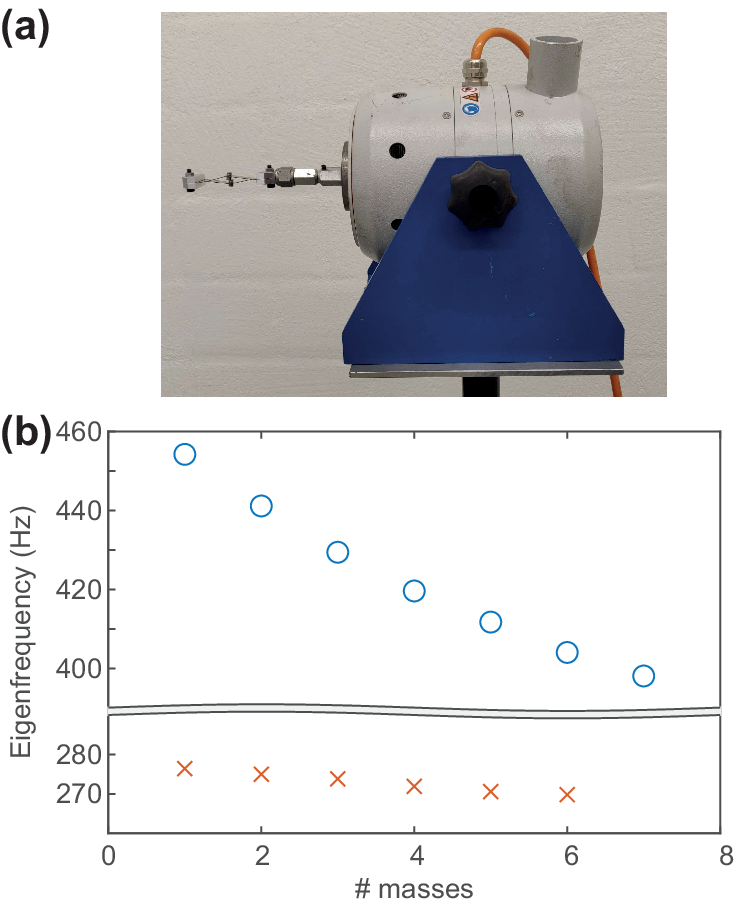}
    \caption{Variation of the first eigenfrequency of TIA 1 and TIA 2 when increasing the amount of amplifying masses.}
    \label{fig:f0}
\end{figure}

The findings of this analytical derivation are now validated by experiments on two TIA and one RIA structure. The three samples are shown in Fig.~\ref{fig:samples}(b). The TIA structures are made out of two plastically bent spring-steel beams. Bolted clamps with a total weight $M = 9.80$~g keep the beams aligned. Beams of the same width and thickness, but with two distinct bending angles were used to investigate the influence of the IA factor. The lengths of TIA1 and TIA2 are $L_1 = 25$~mm and $L_2 = 26$~mm. The static stiffness of both structures was found to be approximately 35~N/mm and 170~N/mm for TIA 1 and 2, respectively~\cite{suppmat}. Based on the measured angles, the amplification factors are $\alpha_1 \approx 5$ and $\alpha_2 \approx 26$. Additional magnetic amplification masses can be added in the beams' bending points as shown in Fig.~\ref{fig:samples}(c). Each individual magnet has a weight of $m = 0.14$~g. It has to be pointed out that the beams' masses ($m_0 = 1.90$~g) also contribute to the amplifying mass. The RIA structure consists of two plexiglass disks with weight $M = 8.78$~g, connected by steel struts. Their angle yields an amplification factor $\alpha_R = 2$.

\begin{figure*}
    \centering
    \includegraphics[width = \textwidth]{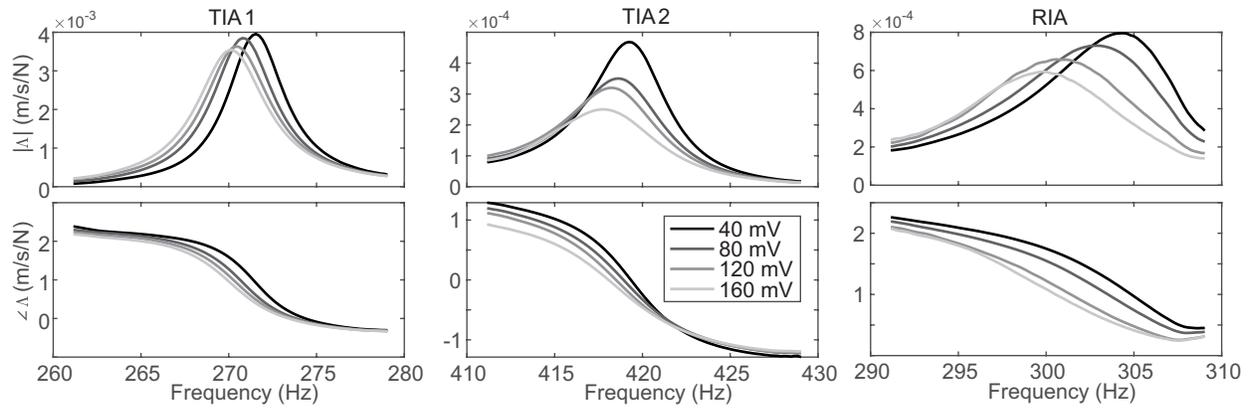}
    \caption{Experimental setup for the FRF measurements of the IA structures (left). Bode plots for the first resonance of 3 IA structures at increasing excitation amplitudes (given in generated voltage amplitude). Four amplifying masses were added to the TIA structures.}
    \label{fig:bode}
\end{figure*}

All structures are excited by an electrodynamic shaker (TIRA TV 51120) connected to one mass, whereas the other mass is allowed to move freely. The input force is measured by a force sensor (PCB 208C01), and the velocity of the free mass is captured simultanously by a laser Doppler vibrometer (Polytec PDV-100). This setup is shown in Fig.\ref{fig:f0}(a). For practical reasons, the simple one-mass system shown in Fig.~\ref{fig:samples}(a) is not feasible for measurements. An infinitely rigid boundary is hard to achieve, and due to the size of the shaker the moving mass is not visible for the laser beam. The resonance behavior of a mass-spring-mass system is recapitulated in \cite{suppmat}. The first eigenfrequency of the resonators is determined by a sine sweep ranging from 100 to 1000~Hz. Subsequently, a 20-Hz range around the eigenfrequency is swept for increasing amplitudes using linear sine sweeps with a duration of 5~s. The spectra of the velocity responses normalized by the input force spectrum are used as FRF resonance curves for further analysis.

First, the effect of the amplification factor is verified by a low-amplitude excitation of TIA 1 and TIA2 for increasing added masses. The amplification factor is difficult to measure exactly: the dynamics of the shaker, the added mass of the force sensor, and a contribution of the proper mass $m_0$ of the elastic beams all change the dynamics of the resonators. However, some qualitative conclusions can be drawn from this experiment. The discrepancy between the eigenfrequencies of TIA1 and 2 is explained by the much lower stiffness of TIA 1. The measured spring constants yield values of the effective mass $\tilde{M} = 2\tilde{K}/\omega^2$ ranging from 22~g to 24~g for TIA1 and 42~g tot 54~g for TIA2. Fig.~\ref{fig:f0}(b) shows that adding 6 magnetic masses to TIA 2 leads to a frequency decrease of over 11\%, whereas the same added mass on TIA 1 only leads to a 2\% shift. The apparent added mass $\Delta m$ for both structures can be estimated from 
\begin{equation}
    \sqrt{\frac{k}{M + \Delta m}} = \frac{f_6}{f_1}\sqrt{\frac{k}{M}},
\end{equation}
where $f_i$ is the eigenfrequency of the system with $i$ added masses. Solving the equation yields $\Delta m_1/M = 0.041$ for TIA~1 and $\Delta m_2/M = 0.263$ for TIA~2, or $\Delta m_2/\Delta m_1 = 6.4$. This relative result agrees fairly well with the theoretical geometric amplification factors ratio $\alpha_2/\alpha_1 = 5.2$. The discrepancy is most probably due to the curved shapes of the structures, changing the effective angles and lengths of the two IA structures.

\begin{figure}
    \centering
    \includegraphics{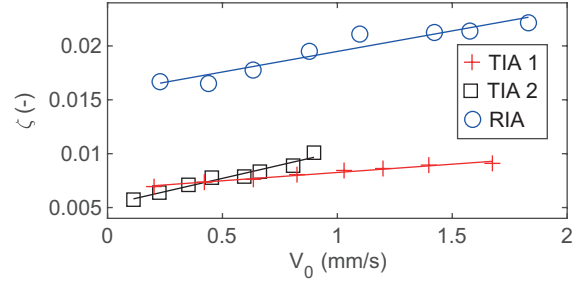}
    \caption{Calculated damping ratio of IA structures as a function of vibration velocity of the mass $M$. The two TIA structures were amplified by 4 added magnetic masses.}
    \label{fig:zeta}
\end{figure}

Now that the ratio of amplification factors has been determined, its influence on the vibration damping can be quantified. To do this, the sine sweeps are repeated 8 times at increasing voltages. The resonance frequency is determined from the maximum of the FRF in the measured frequency range. Fig.~\ref{fig:bode} shows the increasing damping with increasing excitation amplitude for all three resonators. The resonance peaks get wider, the maximum amplitude is reduced, and the slope of the phase diminishes. The maximum slope $\Phi_{max}$ of the measured phase yields the damping ratio according to $\Phi_{max} = -1/(\omega_0\zeta)$. For the TIA samples, the damping ratio is theoretically defined as
\begin{equation}
    \zeta = \frac{8}{\pi L}\frac{m \cos^2\theta_0}{\sin^3\theta_0 \sqrt{k\left(M \sin^2\theta_0 + m\right)}}V_0
\end{equation}
The velocity amplitude $V_0$ is extracted from the amplitude spectrum of the measured time signal. This leads to the measured values of $\zeta$ as a function of the response amplitude shown in Fig.~\ref{fig:zeta}. The results of TIA~1 and TIA~2 both refer to the case of 4 added amplifying masses. The figure clearly reveals the linear increase of the damping ratio with amplitude. Moreover, the proportionality coefficient is higher in the TIA~2 and RIA cases than for the TIA~1 sample. In the case of the TIA resonators, the decrease of $\theta_0$ overpowers the increase of the its stiffness $k$. The RIA resonator is amplified by a much larger mass, which leads to high damping despite the rather large value of $\theta_0$. The nonzero intersection with the vertical axis can be related to the viscoelastic damping of the structure. This appears to be higher for the RIA structure, most probably introduced by the glue connecting the struts with the disks.

\begin{figure}
    \centering
    \includegraphics{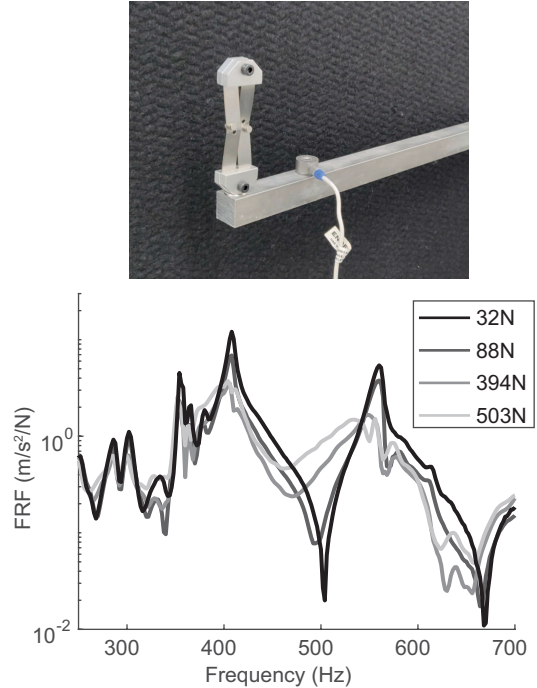}
    \caption{TIA 2 resonator used as tuned mass damper for an aluminum beam (top). FRF of the beam for impact excitations with increasing amplitudes (bottom).}
    \label{fig:tuned}
\end{figure}

IA resonators can be added to vibrating structures as tuned mass dampers, in which case the damping affects the efficiency of the vibration attenuation. The first vibration mode of an aluminum cantilever beam (cross section $18\times18$~mm$^2$) is attenuated by a TIA2 resonator screwed to the beam's free end (Fig.~\ref{fig:tuned}). The beam is excited close to the clamped end using an instrumented hammer (PCB 086B03) using 4 increasing impulse forces. The vibration acceleration close to the resonator is measured by an accelerometer. As expected, the efficiency of the vibration attenuation is reduced for higher impact forces: the antiresonance of the beam is less sharply defined. On the other hand, the secondary resonances left and right of the antiresonance are considerably lower for high impact forces.

In conclusion, we have derived how IA significantly lowers the resonance frequency of spring-mass resonators by adding no (RIA) or a very small (TIA) additional mass, without any assumptions on the coupling angles. However, the kinematic coupling of degrees of freedom in these structures inherently introduces a non-linear term in the equation of motion. The factors contributing to high IA result in a high, amplitude-dependent, damping ratio. This is on the one hand beneficial, since it is a self-controlling system that reduces the mass's displacement at high excitation levels, but it also reduces the vibration reduction efficiency when the resonator is used as a tuned mass damper. This work suggests that special care has to be taken in the design of metamaterials and other structures taking advantage of the IA mechanism. The frequency response might differ considerably from the one achieved by linear models, such as dispersion relations retrieved from unit-cell analysis. However, the damping coefficient of a mechanical resonator is typically hard to control. IA structures have sufficient design parameters to tune both the resonance frequency and its damping, to achieve the optimal tuned damper for a particular application.

\section*{Data Availability}
The data that support the findings of this study are available from the corresponding author upon reasonable request.

\bibliography{IApaper}

\end{document}


\preprint{}

\title[Inherent Non-Linear Damping in Resonators with Inertia Amplification]{Supplementary materials: Inherent Non-Linear Damping in Resonators with Inertia Amplification}
\author{B. Van Damme} 
\email{bart.vandamme@empa.ch.}
\author{G. Hannema}%
\author{L. Sales Souza}%
\author{B. Weisse}
\author{D. Tallarico}%
\author{A. Bergamini}%

\affiliation{ 
Empa, Materials Science and Technology\\
Ueberlandstrasse 129, 8600 Duebendorf\\
Switzerland}%

\date{\today}

\maketitle

\subsection{Static stiffness of the TIA resonators}
\begin{figure}
    \centering
    \includegraphics{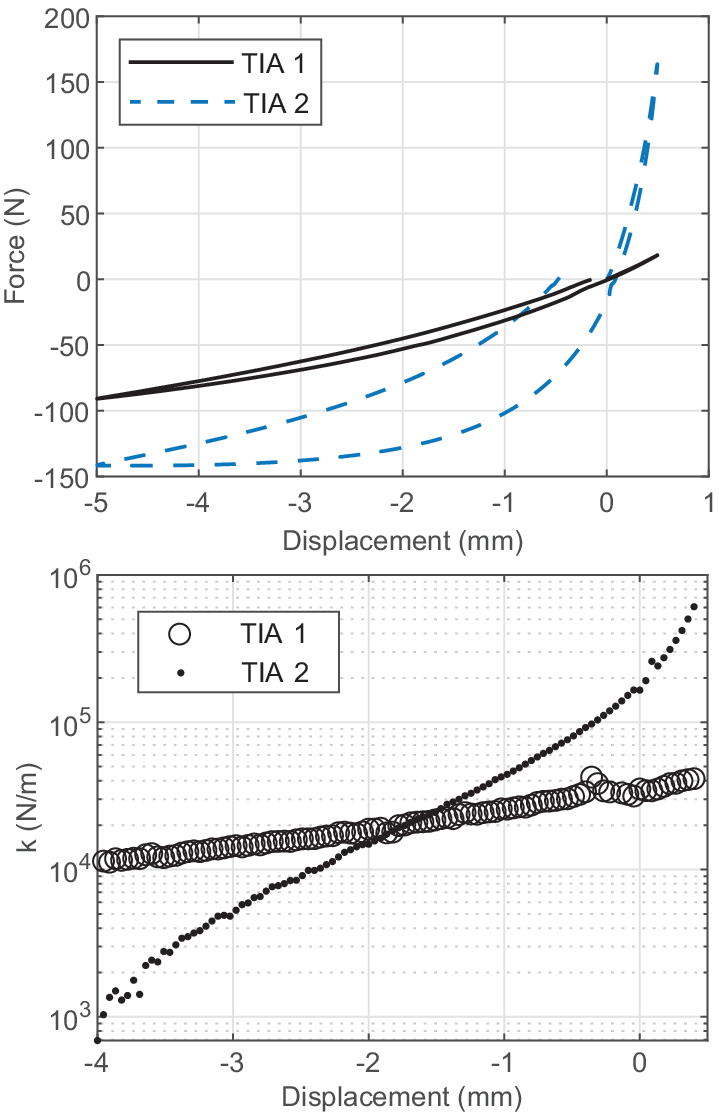}
    \caption{Measured force-displacement diagram (top) and calculated spring constant of the continuous unloading phase (bottom) of the both TIA resonators. The hysteretic loop in the $d-F$ diagram is a result of plastic deformation.}
    \label{fig:static}
\end{figure}
The stiffness under tension and compression of the TIA resonators was tested on  universal test machine of type Zwick Z010 (ZwickRoell GmbH \& Co. KG, Ulm, Germany). The structures are first extended to 0.5~mm, and then compressed to -5.0~mm. The displacement rate is set to 1~mm/min. The force $F$ is measured by the integrated 10 kN load cell (accuracy class 0.5), the displacement $d$ is is measured by the extensometer of type multiXtens II of the test machine of accuracy class 0.5. 7200 force and displacement values are registered every minute to achieve a force-displacement characteristic with high resolution. The stiffness of the structures as a function of the displacement is calculated from the unloading part of the measurement using finite differences. At displacement value $d_i$, the spring constant is
\begin{equation}
    k(d_i) = \frac{F_{i+10} - F_{i-10}}{d_{i+10} - d_{i-10}}.
\end{equation}
The point spacing of 20 samples is used to remove most of the experimental noise. The result shows that the post-buckled beams are much stiffer during extension than during compression. Plastic deformation close to the clamps and in the bending point of the beams is another sorce of nonlinearity, resulting in a hysteretic force-deformation relation. 

For the small displacement amplitudes considered in the dynamic experiments, less than 1~$\mu$m, we assume that the spring constant remains unchanged at its equilibrium value $d = 0$.

\subsection{Equation of motion of RIA systems}\label{app:RIA}
\begin{figure}
    \centering
    \includegraphics[width = \columnwidth]{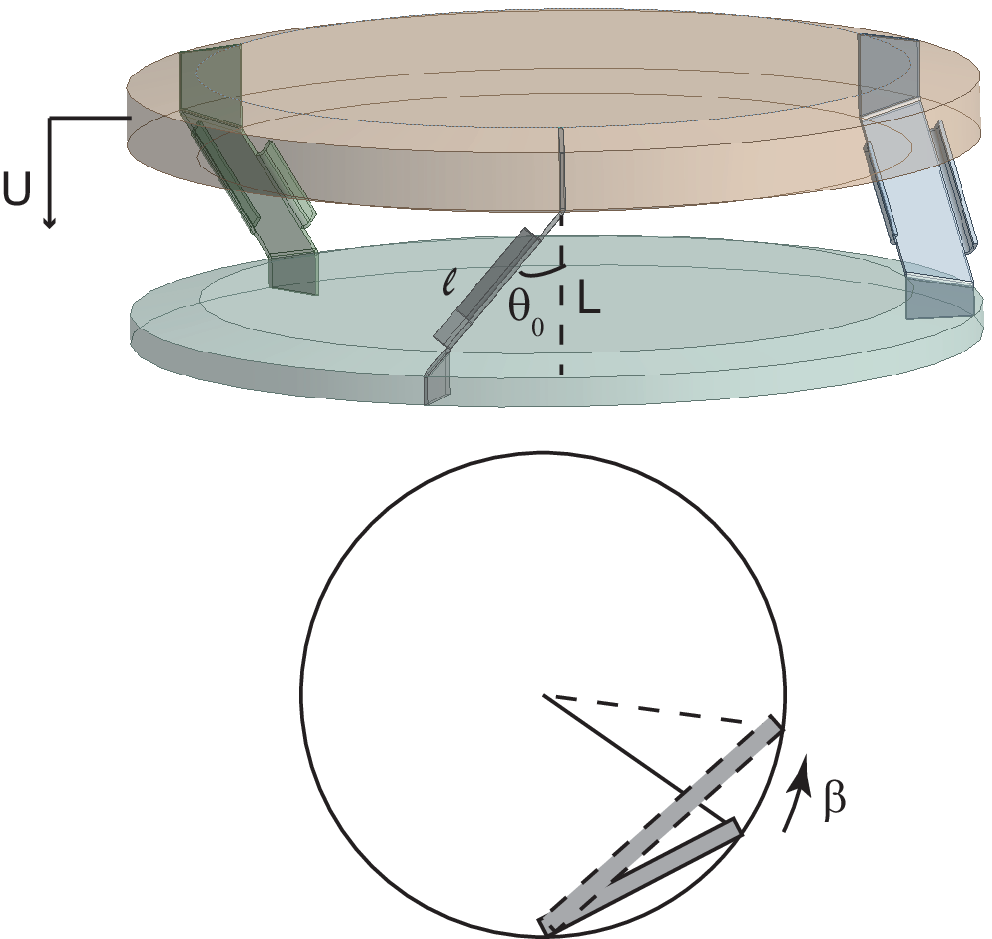}
    \caption{Schematic representation of a RIA structure. The translation $U$ leads to a disk rotation $\beta$.}
    \label{fig:RIA_motion}
\end{figure}
The coupling between the translation $U$ of the top disk with its rotation can be clarified from the schematic representation of Fig.~\ref{fig:RIA_motion}. The connecting strut is shown before and after displacement, showing that its projected length becomes larger. The chords before and after deformation have lengths $\sqrt{\ell^2 - L^2}$ and $\sqrt{\ell^2 - (L - U)^2}$, respectively. Without too much error, chords' lengths equal the arch lengths, and the rotation angle is then given by
\begin{equation}
    \beta = \frac{\sqrt{\ell^2 - (L - U)^2} - \sqrt{\ell^2 - L^2}}{R},
\end{equation}
$R$ being the radius of the disk. The angular velocity results from the derivative with respect to time: 
\begin{equation}
    \dot{\beta} = \frac{L - U}{\sqrt{\ell^2 - (L - U)^2}}\frac{\dot{U}}{R}.
\end{equation}
Knowing that $\ell = L/\cos \theta_0$, this result is very similar to the one of TIA structures. The kinetic energy required for the Lagrangian is given by $T = M\dot{U}^2/2 + I\dot{\beta}^2/2$, where $I = MR^2/2$ is the moment of inertia of the rotating disk and $M$ its mass. Deriving the equation of motion in a similar way as for the TIA structure yields 
\begin{eqnarray}
\tilde{M} &=& M + \frac{M}{2}\left( 1 + \frac{(1-\epsilon)^2}{A} \right)\\
\tilde{C} &=& \frac{M}{2L} \left( \frac{1-\epsilon}{A}  + \frac{(1-\epsilon)^3}{A^2}\right)\\
\tilde{K} &=& k \\
A &=& \frac{1}{\cos^2 \theta_0} - (1-\epsilon)^2,
\end{eqnarray}
using the strain $\epsilon = U/L$. The results are very similar to the TIA equations, except that the amplification now arises through the main mass $M$. 

\subsection{Resonance and damping ratio of a mass-spring-mass system}\label{app1}
Two identical masses $M$ connected by a spring with spring constant $k$ and damper with damping coefficient $c$ are excited by a force on one mass, whereas the other mass is kept free. Their displacements $u_1$ and $u_2$ are described by the equations of motion
\begin{eqnarray}
F &=& M \ddot{u}_1 + c(\dot{u}_1 - \dot{u}_2) + k(u_1 - u_2)\\
0 &=& M \ddot{u}_2 + c(\dot{u}_2 - \dot{u}_1) + k(u_2 - u_1).
\end{eqnarray}
For this linear case, the complex response $u_j = U_j \exp(i \omega t)$ ($j=1,2$) to a harmonic force $F = F_0 \exp(i \omega t)$ can be used, leading to the frequency response function of the free mass
\begin{equation}\label{eq:resonance}
    \Lambda = \frac{U_2}{F_0} = \frac{M \omega^2 - i\omega c - k}{M\omega^2(M\omega^2 - 2i\omega c - 2k)}.
\end{equation}
Apart from a rigid-body resonance at $\omega = 0$, the non-trivial resonance frequency for the non-damped case is given by $\omega_0 = \sqrt{2k/M}$. The damped resonance frequency occurs at $\omega_d = \sqrt{2Mk - 2c^2}/M$, which can be rewritten as $\omega_d = \omega_0\sqrt{1-\zeta^2}$, defining the damping ratio $\zeta = c/\sqrt{kM}$.

The damping ratio can be derived from the measured FRF in two ways: from the peak width or from the slope of the phase angle at the resonance frequency. In this work, we choose for the second approach. During the sweep excitation the amplitude might vary, which indirectly results in a frequency-dependent damping. Therefore, the resonance peak is deformed with respect to the constant-damping formula in eq.~\ref{eq:resonance}. The maximum slope of the phase shift depends only on the response amplitude at $\omega = \omega_d$ and is thus a more reliable estimator. The phase shift $\phi$ is given by 
\begin{equation}
    \angle \Lambda = \arctan\left(\frac{Im(\Lambda)}{Re(\Lambda)}\right).
\end{equation}
Substituting the value for $\omega_d$ into the derivative of $\phi$ with respect to $\omega$ yields
\begin{equation}
    \Phi_{max} = \left.\frac{d\phi}{d \omega}\right|_{\omega_d} = -\frac{1}{\omega_0 \zeta}\frac{1 + \zeta^2 - 2\zeta^4}{1 + 2\zeta^2 - 2\zeta^4}, 
\end{equation}
which can be approximated by 
\begin{equation}\label{eq:zeta}
    \Phi_{max} = -1/(\omega_0\zeta)
\end{equation} 
for small values of $\zeta$. Notice that the formulas for the damped resonance frequency and the estimation of the damping ratio are identical to those for the well-known spring-mass-damper system, except that in that case the damping ratio is defined as $\zeta = c/2\sqrt{kM}$.